\documentclass[12pt]{article}
\usepackage{amsmath}
\usepackage{graphicx}
\usepackage{natbib}
\usepackage{url} 

\newcommand{\blind}{0}

\usepackage{multicol}
\usepackage[inline]{enumitem} 
\usepackage{float}
\usepackage{bm}
\usepackage{amsthm}
\usepackage{amssymb}
\usepackage{comment}
\usepackage{enumerate}
\usepackage{algorithm}
\usepackage{algorithmicx}\usepackage{algpseudocode}
\usepackage{latexsym}
\usepackage{multirow}\usepackage{bm}%
\usepackage{enumitem}

\newtheorem{assumption}{Assumption}

\newtheorem{proposition}{Proposition}
\newtheorem{theorem}{Theorem}

\newtheorem{lemma}{Lemma}

\def\independenT#1#2{\mathrel{\rlap{$#1#2$}\mkern2mu{#1#2}}}

\usepackage[compact]{titlesec}
\allowdisplaybreaks
\newcommand{\spacingset}[1]{\renewcommand{\baselinestretch}%
{#1}\small\normalsize}

\newcommand{\bone}{\mathbf{1}}

\newcommand{\logit}{\text{logit}}

\newcommand{\pr}{\mathbb{P}}

\newcommand{\aipw}{\mathrm{aw}}

\newcommand{\rct}{\mathcal{I}_n}
\newcommand{\rwe}{\mathcal{I}_m}
\newcommand{\expect}{\mathbb{E}}
\newcommand{\independent}{\protect\mathpalette{\protect\independenT}{\perp}}
\newcommand{\ri}{\right}
\newcommand{\lf}{\left}
\newcommand{\be}{\eta}
\newcommand{\bx}{\boldsymbol{x}}
\newcommand{\bX}{\boldsymbol{X}}

\newcommand{\Leps}{L^{\zeta}}
\newcommand{\thetaw}{\hat{\theta}_{w}}
\newcommand{\transpose}{\intercal}

\newcommand{\define}{\overset{\mathrm{\Delta}}{=}}
\newcommand{\indicator}{\bone}

\bibpunct{(}{)}{;}{autheryear}{,}{;}

\begin{document}

\def\spacingset#1{\renewcommand{\baselinestretch}%
{#1}\small\normalsize} \spacingset{1}


\if0\blind
{
  \title{\bf Transfer learning of individualized treatment rules from experimental
to real-world data}
  \author{Lili Wu and Shu Yang\thanks{
    The authors gratefully acknowledge \textit{NSF grant DMS 1811245, NCI grant P01 CA142538, NIA grant 1R01AG066883, and NIEHS grant 1R01ES031651.}}\hspace{.2cm}\\
    Department of Statistics, North Carolina State University}
  \maketitle
} \fi

\if1\blind
{
  \bigskip
  \bigskip
  \bigskip
  \begin{center}
    {\LARGE\bf Title}
\end{center}
  \medskip
} \fi

\bigskip
\begin{abstract}
Individualized treatment effect lies at the heart of precision medicine.
Interpretable individualized treatment rules (ITRs) are desirable
for clinicians or policymakers due to their intuitive appeal and transparency.
The gold-standard approach to estimating the ITRs is randomized experiments,
where subjects are randomized to different treatment groups and the
bias is minimized to the extent possible. However, experimental data
are limited in external validity because of their selection restrictions
and therefore are not representative of the target real-world population.
Conventional learning methods of optimal interpretable ITRs for a
target population based only on experimental data are biased. On the
other hand, real-world data (RWD) are becoming popular and provide
a representative sample of the population. To learn the generalizable optimal interpretable ITRs, we propose
an integrative transfer learning method based on weighting schemes to calibrate the covariate distribution of the experiment to that of the RWD.
We show that the proposed ITR estimator has a theoretical
guarantee of the risk consistency. We evaluate the transfer learner
based on the finite-sample performance through simulation and apply
it to a real data application of a job training program.
\end{abstract}

\noindent%
\vfill

\newpage
\spacingset{1.5} 
\section{Introduction}
\label{sec:intro}

A personalized recommendation, tailored to individual characteristics,
is becoming increasingly popular in real life such as healthcare,
education, e-commerce, etc. An optimal \textit{individualized treatment
rule} (ITR) is defined as maximizing \textit{the mean value function
}of an outcome of interest over the target population if all individuals
in the whole population follows the given rule. Many machine learning
approaches to estimating the optimal ITR are available, such as outcome-weighted
learning with support vector machine \citep{zhao2012estimating,zhou2017residual},
regression-based methods with Adaboost \citep{kang2014combining},
regularized linear basis function \citep{qian2011performance}, K-nearest
neighbor \citep{zhou2017causal} and generalized additive model \citep{moodie2014q}.
However, the ITRs derived by machine learning methods can be too complex
to extract practically meaningful insights for policymakers. For example,
in healthcare, clinicians need to scrutinize the estimated ITRs for
scientific validity but the black-box rules conceal the clear relationship
between patient characteristics and treatment recommendations. In
these cases, parsimonious and interpretable ITRs are desirable.

It is well known that randomized experiments are the gold standard
for learning the optimal ITRs because randomization of treatments
in the design stage can ensure no unmeasured confounding \citep{greenland1990randomization}.
However, randomized experiments might lack external validity or representativeness
\citep{rothwell2005external} because of the selective criteria for
individual eligibility in the experiments. Consequently, the distribution
of the covariates in the randomized experiments might differ from
that in the target population. Thus, a parsimonious ITR constructed
from randomized experiments cannot directly generalize to a target
population \citep{zhao2019robustifying}. Alternatively, real-world
data (RWD) such as population survey data usually involve a large
and representative sample to the target population. A large body of
literature in statistics has focused on transporting or generalizing
the average treatment effects from the experiments to a target population
or a different study population \citep[e.g.,][]{cole2010generalizing,hartman2015sample}.
One exception for the ITRs is \citet{zhao2019robustifying, mo2020learning} which
estimates the optimal linear ITR from the experimental data by optimizing
the worst-case quality assessment among all covariate distributions
in the target population satisfying some moment conditions or distributional closeness, while
we address the problem in a different situation where we already have
a representative sample at hand. 

In this article, we propose a general framework of estimating a generalizable
interpretable optimal ITR from randomized experiments to a target
population represented by the RWD. We propose\textit{ transfer weighting}
that shifts the covariate distribution in the randomized experiments
to that in the RWD. To correctly estimate the population optimal ITR,
transfer weights are used in two places. First, the empirical estimator
of the population value function is biased when considering only the
experimental sample, and hence the transfer weights are used to correct
the bias of the estimator of the value function. Second, the nuisance
functions in the value function such as the propensity score and the
outcome mean also require transfer weighting in order to estimate
the population counterparts. Then we can obtain the weighted interpretable
optimal ITR by reformulating the problem of maximizing the weighted
value to that of minimizing the weighted classification error with
flexible loss function choices. We consider different parametric and
nonparametric approaches to learning transfer weights. Moreover, a
practical issue arises regarding the choice of transfer weights. Weighting
can correct for the bias but also reduce the effective sample size
and increase the variance of the weighted value estimator. To address
the bias-variance tradeoff, we propose a cross-validation procedure
that maximizes the estimated population value function to choose among
the weighting schemes. Lastly, we provide the theoretical guarantee
of the risk consistency of the proposed transfer weighted ITRs.

The remaining of the article is organized as follows. We introduce
our proposed method in Section \ref{sec:meth}. In Section \ref{sec:asy},
we study the asymptotic properties of the proposed estimator to guarantee
its performance in the target population theoretically. In Section
\ref{sec:sim}, we use various simulation settings to illustrate how
this approach works. In Section \ref{sec:application}, we apply the
proposed method to the National Supported Work program \citep{lalonde1986evaluating}
for improving the decisions of whether to recommend the job training
program based on the individual characteristics over the target population.
We conclude the article in Section \ref{sec:Discussion}.

\section{Basic Setup}

\subsection{Notation}

Let $\bX\in\mathcal{X}\subseteq\mathbb{R}^{p}$ be a vector of pre-treatment
covariates, $A\in\mathcal{A}=\{0,1\}$ the binary treatment, and $Y\in\mathbb{R}$
the outcome of interest. Under the potential outcomes framework \citep{rubin1974estimating},
let $Y^{*}(a)$ denote the potential outcome had the individual received
treatment $a$ $\in\mathcal{A}$. Define a treatment regime as a mapping
$d:\mathcal{X}\rightarrow\mathcal{A}$ from the space of individual
characteristic to the treatment. Then we define the potential outcome
under regime $d$ as $Y^{*}(d)=\sum_{a\in\mathcal{A}}Y^{*}(a)\indicator\{d(\bX)=a\}$,
the value of regime $d$ as $V(d)=\mathbb{E}\{Y^{*}(d)\}$, where
the expectation is taken over the distribution in the target population,
and the optimal regime $d^{\text{opt}}$ as the one satisfying $V(d^{\text{opt}})\geq V(d),\:\forall d.$
As discussed in the introduction, in many applications such as healthcare,
interpretable treatment regimes are desirable since they might bring
more domain insights and be more trustworthy to clinicians compared
with the complex black-box regimes. Thus, it is valuable to learn
an optimal interpretable regime. Specifically, we focus on the regimes
with linear forms, i.e, $d(\bX;\be)=\indicator(\eta_{0}+\eta_{1}^{\intercal}\bX>0),$
denoted by $d_{\eta}$ for simplicity. 

We aim at learning ITRs which lead to the highest benefits for the
target population with size $N$, $\{\bX_{i},Y_{i}^{*}(0),Y_{i}^{*}(1)\}_{i=1}^{N}$.
Suppose we have access to two independent data sources: the experiment
data with $\{(\bX_{i},A_{i},Y_{i})\}_{i\in\rct}$ and the RWD with
$\{\bX_{i}\}_{i\in\rwe}$, where $\rct,\rwe\subseteq\{1,\dots,N\}$
are samples' index sets with size $n,m$, respectively. In the experimental
data, individuals are selected based on certain criteria and then
are completely randomized to receive treatments. The experimental
data might induce selection bias which leads to a different distribution
of $\bX$ in experiments from that in the target population. On the
other hand, we assume that individuals in the RWD are a random sample
from the target population, thus the distribution of the covariates
in the RWD can characterize that of the target population. Let $S_{i}$
be the binary indicator of the $i$th individual participates in the
randomized experiment: $S_{i}=1$ if $i\in\mathcal{I}_{n}$ and $0$
if $i\in\mathcal{I}_{m}$. We denote $\pi_{A}(\bX)$ as the propensity
score of receiving the active treatment $P(A=1\mid\bX)$, $Q(\bX,A)$
as the conditional expectation of outcomes given covariates and treatments
$\expect(Y\mid\bX,A)$, and $\tau(\bX)$ as the contrast function $Q(\bX,1)-Q(\bX,0)$. 

We make the following assumptions, which are standard and well-studied
assumptions in the causal inference and treatment regime literature
\citep{chakraborty2013statistical}.

\begin{assumption}
\begin{enumerate*}[label={\alph*:}, ref={Assumption~\theassumption.\alph*}]
  \item \label{asu: nuc}  $A\independent{\{Y^{*}(0),Y^{*}(1)\}}\mid(\bX,S=1)$. 
  \item \label{asu: consistency}$Y=Y^{*}(A)$. 
  \item  \label{asu: positivity}There exist $c_{1},c_{2}\in(0,1)$
such that $\pi_{A}(\bX)\in[c_{1},c_{2}]$ for all $\bX\in\mathcal{X}.$
\end{enumerate*}
\end{assumption}

\subsection{Existing learning methods for ITRs \label{subsec:Existing-learning-methods}}

Identifying the optimal individualized treatment rules has been studied
in a large body of statistical literature. One category is regression-based
by directly estimating the conditional expectation of the outcome
given patients' characteristics and treatments received \citep{murphy2005generalization}. Both parametric and nonparametric models can be used to approximate
$Q(\bX,A)$. To encourage the interpretability, one strategy is to
posit a parametric model as $Q(\bX,A;\beta)$, and the estimated optimal
ITR within the class indexed by $\beta$ is defined as $\indicator\{\tau(\bX;\hat{\beta})>0\}$.

Another category is value-based by first estimating the value of an
ITRs and then searching the ITR that renders the highest value. One
strategy is to estimate $\pi_{A}(\bX)$ first, where one common way
is to posit a parametric model as $\pi_{A}(\bX_{i};\gamma)$ such
as logistic regression or a constant, and then use inverse probability
weighting (IPW; \citealp{horvitz1952generalization}) to estimate
the value of an ITR indexed by $\be$. This estimator is consistent to the true value only if $\pi_{A}(\bX_{i};\gamma)$
is correctly specified and can be unstable if the estimate $\pi_{A}(\bX_{i};\hat{\gamma})$
is close to zero. The augmented inverse probability weighting (AIPW; \cite{zhang2012robust}) approach is introduced
to combine IPW and outcome regression estimator $\widehat{Q}(\bX_i, A)$ to
estimate the value 

\begin{equation}
\begin{split}
\widehat{V}_{\text{aw}}(d_{\be};\mathcal{I}_{n})=\frac{1}{n}\sum_{i\in\rct}\Bigg(&\left[\frac{A_{i}d(\bX_{i};\be)}{\pi_{A}(\bX_{i};\hat{\gamma})}+\frac{(1-A_{i})\left\{ 1-d(\bX_{i};\be)\right\} }{1-\pi_{A}(\bX_{i};\hat{\gamma})}\right]\left[Y_{i}-\widehat{Q}\left\{ \bX_{i},d(\bX_{i};\be)\right\} \right]\\
&-\widehat{Q}\left\{ \bX_{i},d(\bX_{i};\be)\right\}\Bigg). \label{eq:awV}
\end{split}
\end{equation}
 
The AIPW estimator is doubly robust in the sense that it will be consistent
to the value of the ITR $d_{\be}$ if either the model for $\pi_{A}(\bX)$
or $Q(\bX,A)$ is correctly specified, but not necessarily both. 

Learning the optimal ITR by optimizing the value can also be formulated
as a classification problem \citep{zhang2012estimating}. The classification
perspective has equivalent forms to all the regression-based and value-based
methods above, and can also bring us additional optimization advantages
over directly value searching. 

\begin{lemma}\label{lemma:1} The optimal treatment rule by maximizing
$V(d_{\be})$ in $\eta$ is equivalent to that by minimizing the risk
\begin{equation}
\expect\left(\big|\tau(\bX)\big|\Big[\indicator\big\{\tau(\bX)>0\big\}-d(\bX;\be)\Big]^{2}\right)\label{eq:classificationObj}
\end{equation}
in $\eta$ \citep{zhang2012estimating}.

\end{lemma}

Lemma \ref{lemma:1} reformulates the problem of estimating an optimal
treatment rule as a weighted classification problem, where $\mid\tau(\bX)\mid$
is regarded as the weight, $\indicator\{\tau(\bX)>0\}$ the binary
label, and $d(\bX;\eta)$ the classification rule of interest. Let
$l_{0-1}$ be a 0-1 loss, i.e., $l_{0-1}(u)=\indicator(u\leq0)$.
The objective function (\ref{eq:classificationObj}) can also be rewritten as
\begin{equation}
\resizebox{.94\hsize}{!}{$\mathcal{R}_{\mathcal{F}}(\tau;\be,l_{0-1})=\expect\left\{ \mathcal{\mathcal{L}}(\tau;\be,l_{0-1})\right\} ,\ \ \mathcal{\mathcal{L}}(\tau;\be,l_{0-1})=\mid\tau(\bX)\mid l_{0-1}\left\{ \left[2\indicator\{\tau(\bX)>0\}-1\right]f(\bX;\be)\right\},\label{eq:obj_0_1_loss}$}
\end{equation}
where $f(\bx;\be)=\be^{\intercal}\bx$, $d(\bx;\be)=\indicator\lf\{f(\bx;\be)>0\ri\}$,
$\mathcal{F}$ is a certain class covering the parameters of the ITRs,
i.e., $\be\in\mathcal{F}$. 

Let $\hat{\tau}_{i}$ be an estimator of $\tau(\bX_{i})$. From Lemma
\ref{lemma:1}, to estimate the optimal ITR, one can minimize the
following empirical objective function:
\begin{eqnarray}
n^{-1}\sum_{i\in\rct}\mid\hat{\tau}_{i}\mid\{\indicator{(\hat{\tau}_{i}>0)}-d(\bX_{i};\be)\}^{2} & = & n^{-1}\sum_{i\in\rct}\mid\hat{\tau}_{i}\mid\indicator\left\{ \indicator\left(\hat{\tau}_{i}>0\right)\neq d(\bX_{i};\be)\right\} \nonumber \\
 & = & n^{-1}\sum_{i\in\rct}\mid\hat{\tau}_{i}\mid l_{0-1}\left[\left\{ 2\indicator(\hat{\tau}_{i}>0)-1\right\} f(\bX_{i};\be)\right]\nonumber \\
 & = & n^{-1}\sum_{i\in\rct}\mathcal{L}(\hat{\tau}_{i};\be,l_{0-1}).\label{eq:empObj}
\end{eqnarray}
In this classification framework, we can also consider the three different
approaches to estimating $\tau(\bX_{i})$, which have the equivalent
forms and enjoy the same properties with the regression- and value-based
estimator. For example, The AIPW estimator of $\tau(\bX_{i})$ is
\[
\hat{\tau}_{\aipw,i}=\left[\frac{\indicator(A_{i}=1)\big\{ Y_{i}-\widehat{Q}(\bX_{i},1)\big\}}{\hat{\pi}_{A}(\bX_{i})}+\widehat{Q}(\bX_{i},1)\right]-\left[\frac{\indicator(A_{i}=0)\big\{ Y_{i}-\widehat{Q}(\bX_{i},0)\big\}}{1-\hat{\pi}_{A}(\bX_{i})}+\widehat{Q}(\bX_{i},0)\right].
\]
The solution obtained by minimizing (\ref{eq:empObj}) in $\eta$
with replacing $\hat{\tau}_{i}$ with $\hat{\tau}_{\text{aw},i}$
is equivalent to that obtained by maximizing the AIPW value estimator
$\widehat{V}_{\text{aw}}(d_{\be};\rct)$ in (\ref{eq:awV}) and enjoys
the doubly robust property.

When the experimental sample is representative of the target population,
the above ITR learning methods can estimate the population optimal
ITRs that leads to the highest value. To correct for the selection
bias of the experimental sample, we propose an integrative strategy
that leverages the randomization in the experimental sample and the
representativeness of the RWD. As pointed out by \citet{zhao2019robustifying},
the optimal rule obtained from the experimental sample is also optimal
for the target population if no restriction to the class of $d,$
but such optimal rules are too complex. Therefore, we focus on learning
interpretable and parsimonious ITRs and thus integrative methods are
required to correct for the selection bias of the experimental sample. 

\section{Integrative transfer learning of ITRs\label{sec:meth}}

\subsection{Assumptions for identifying population optimal ITRs \label{sec:rule} }

To identify population optimal ITRs, we assume the covariates $\bX$
can fully explain the selection mechanism of inclusion into the randomized
experiment and the covariate distribution in the RWD is representative
to the target population. 

\begin{assumption}
\begin{enumerate*}[label={\alph*:}, ref={Assumption~\theassumption.\alph*}]
  \item \label{asu:cond indep} $S\independent\left\{ Y^{*}(0),Y^{*}(1),A\right\} \mid\bX$.
  \item \label{asu:unbias}$f(\bX\mid S=0)=f(\bX)$.
  \item \label{asu:positivity_pi_S} There exist $c_{3},c_{4}\in(0,1)$ such that $\pi_{S}(\bX)\in[c_{3},c_{4}]$ for all $\bX\in\mathcal{X}.$
\end{enumerate*}
\end{assumption}

Combining \ref{asu:cond indep} and \ref{asu: consistency}
implies that the experimental sample is non-informative, i.e., the
sampling score is $P(S=1\mid A,\bX,Y)=P(S=1\mid\bX)=\pi_{S}(\bX)$.
It requires that the observed covariates $\bX$ is rich enough such
that it can decide whether or not the subject is selected into the
experimental data.

\ref{asu:unbias} requires that covariate distribution
in the RWD is the same as that in the target population so that we
can leverage the representativeness of the RWD. As a result, it implies
that $m^{-1}\sum_{i\in\mathcal{I}_{m}}g(\bX_{i})$ would be an unbiased
estimator of $\expect{\{g(\bX)\}}$, the covariate expectation in
the target population for any $g$. \ref{asu:cond indep} and \ref{asu:unbias} are also adopted
in the covariate shift problem in machine learning, a sub-area in
domain adaptation, where standard classifiers do not perform well
when the independent variables in the training data have a different
covariate distribution from that in the testing data. Re-weighting
individuals to correct the over- or under-representation is one way
to cope with the covariate shift in classification problems \citep{kouw2018introduction}. 

\ref{asu:positivity_pi_S} requires each subject has chance
to be selected into the experimental data so as to guarantee the existence
of $\pi_{S}^{-1}(\bX)$.

The following proposition identifies the population value function
using the inverse of sampling score as the transfer weight, and leads to an equivalent weighted classification
loss function to identify the population optimal ITR.

\begin{proposition} \label{prop:1} Let the transfer weight
be $w=\pi_{S}^{-1}(\bX)$, under \ref{asu: consistency},
\ref{asu:cond indep} and \ref{asu:positivity_pi_S}, the population
value of an ITR $d_{\be}$ is identified by $V(d_{\be})=\expect\left[\omega SY^{*}\left\{ d_{\be}(\bX)\right\} \right]$. Then the population optimal ITR can
be obtained by minimizing 
\[
\expect\left(wS\big|\tau(\bX)\big|\Big[\indicator\big\{\tau(\bX)>0\big\}-d_\be(\bX)\Big]^{2}\right) \text{ in $\eta$.}
\]
\end{proposition}

Two challenges arise for using Proposition \ref{prop:1} to estimate
the optimal ITR: (i) the weights $w$ are unknown and (ii) the estimators
of $\tau(\bX)$ discussed so far such as $\hat{\tau}_{\text{aw}}$ are biased
based only on the experimental sample. In the following subsections,
we present the estimation of the transfer weights and unbiased estimators
of $\tau(\bX)$. 

\subsection{Estimation of transfer weights}

We consider various parametric and nonparametric methods to estimate
the transfer weights $w$. Each method has its own benefits: parametric
methods are easy to implement and efficient if the models are correctly
specified, while nonparametric methods are more robust and less sensitive
to model misspecification.

\subsubsection{Parametric approach \label{subsec:Parametric-approach}}

For parametric approaches, we assume the target population size $N$
is known. Similar to modeling propensity score $\pi_{A}(\bX)$, we
can posit a logistic regression model for $\pi_{S}(\bX)$; i.e., $\logit\{\pi_{S}(\bX;\alpha)\}=\alpha_{0}+\alpha_{1}^{\intercal}\bX$,
where $\alpha=(\alpha_{0},\alpha_{1}^{\intercal})^{\intercal}\in\mathbb{R}^{p+1}$.
The typical maximum likelihood estimator of $\alpha$ can be obtained
by 
\begin{align}
\Hat{\alpha}= & \underset{\alpha}{\mathrm{argmax}}\frac{1}{N}\sum_{i=1}^{N}\left[S_{i}\log\pi_{S}(\bX_{i};\alpha)+(1-S_{i})\log\left\{ 1-\pi_{S}(\bX_{i};\alpha)\right\} \right]\nonumber \\
= & \underset{\alpha}{\mathrm{argmax}}\frac{1}{N}\sum_{i=1}^{N}\left[S_{i}(\alpha_{0}+\alpha_{1}^{\intercal}\bX_{i})-\mathrm{log}\left\{ 1+\mathrm{exp}(\alpha_{0}+\alpha_{1}^{\intercal}\bX_{i})\right\} \right].\label{eq:mle}
\end{align}
However, the second term $N^{-1}\sum_{i=1}^{N}\mathrm{log}\left\{ 1+\mathrm{exp}(\alpha_{0}+\alpha_{1}^{\intercal}\bX_{i})\right\} $
in (\ref{eq:mle}) is not feasible to calculate because we can not
observe $\bX$ for all individuals in the population. The key insight
is that this term can be estimated by $m^{-1}\sum_{i\in\rwe}\mathrm{log}\left\{ 1+\mathrm{exp}(\alpha_{0}+\alpha_{1}^{\intercal}\bX_{i})\right\} $
based on the RWD. This strategy leads to our modified maximum likelihood
estimator
\[
\Hat{\alpha}=\underset{\alpha}{\mathrm{argmax}}\left[\frac{1}{N}\sum_{i=1}^{N}S_{i}(\alpha_{0}+\alpha_{1}^{\intercal}\bX_{i})-\frac{1}{m}\sum_{i=1}^{m}\mathrm{log}\left\{ 1+\mathrm{exp}(\alpha_{0}+\alpha_{1}^{\intercal}\bX_{i})\right\} \right].
\]

Alternatively, \ref{asu:unbias} leads to unbiased estimating
equations of $\alpha$ 

\[
\expect\left\{ \frac{Sg(\bX)}{\pi_{S}(\bX;\alpha)}\right\} =\expect\{g(\bX)\},
\]
for any $g$. Therefore, we can use the following estimating equations 

\[
\frac{1}{N}\sum_{i=1}^{N}\frac{S_{i}g(\bX_{i})}{\pi_{S}(\bX_{i};\alpha)}=\frac{1}{m}\sum_{i\in\rwe}g(\bX_{i})
\]
for $\alpha$, where $g(\bX)\in\mathbb{R}^{p+1}$. For simplicity,
one can take $g(\bX)$ as $\partial\logit\{\pi_{S}(\bX;\alpha)\}/\partial\alpha$. 

\subsubsection{Nonparametric approach\label{subsec:Nonparametric-approach}}

The above parametric approaches require $\pi_{S}(\bX;\alpha)$ to
be correctly specified and the population size $N$ to be known. These
requirements may be stringent, because the sampling mechanism into
the randomized experiment is unknown and the population size is difficult
to obtain. We now consider the constrained optimization algorithm
of \citet{wang2020minimal} to estimate the transfer weights by
\begin{align}
 & \underset{w}{\text{min}}\ \sum_{i=1}^{N}S_{i}w_{i}\log w_{i}\nonumber \\
 & \text{subject to}\ \Big|\sum_{i=1}^{N}w_{i}S_{i}g_{k}(\bX_{i})-\frac{1}{m}\sum_{i\in\rwe}g_{k}(\bX_{i})\Big|\leq\sigma_{k}\ \:\:(k=1,\dots,K),\label{eq: minimial weight}
\end{align}
where $\sigma_{k}\geq0;w_{i}\in[0,1],\:i\in\mathcal{I}_{n},$ $\sum_{i\in\mathcal{I}_{n}}w_{i}=1,$
$\text{ and }\{g_{1},\ldots,g_{K}\}$ can be chosen as the first-,
second- and higher order moments of the covariate distribution. The
constants $\sigma_{k}$ are the tolerate limits of the imbalances
in $g_{k}$. When all the $\sigma_{k}$'s are taken 0, (\ref{eq: minimial weight})
becomes the exact balance, and the solution $w$ becomes the entropy
balancing weights \citep{hainmueller2012entropy}. The choice of $\sigma_{k}$ is related to the standard bias-variance
tradeoff. On the one hand, if $\sigma_{k}$ is too small, the weight
distribution can have a large variability in order to satisfy the
stringent constraints, and in some extreme scenarios the weights may
not exist. On the other hand, if $\sigma_{k}$ is too large, the covariate
imbalances remain and therefore the resulting weights are not sufficient
to correct for the selection bias. Following \citet{wang2020minimal},
we choose $\sigma_{k}$ from a pre-specified set by the tuning algorithm described in Supplementary Materials.

\subsection{Estimation of $\tau(\bX)$}

The transfer weights are also required to estimate $\tau(\bX)$ from
the experimental data. In parallel to Section \ref{subsec:Existing-learning-methods},
we can obtain the weighted estimators of $\tau(\bX).$ For example, the weighted AIPW estimator of $\tau(\bX_{i})$
\[
\resizebox{0.98\hsize}{!}{$
\hat{\tau}_{\aipw,i}^{w}=\Big[\frac{\indicator(A_{i}=1)\big\{ Y_{i}-\widehat{Q}_{w}(\bX_{i},1)\big\}}{\hat{\pi}_{A}^{w}(\bX_{i})}+\widehat{Q}_{w}(\bX_{i},1)\Big]-\Big[\frac{\indicator(A_{i}=0)\big\{ Y_{i}-\widehat{Q}_{w}(\bX_{i},0)\big\}}{1-\hat{\pi}_{A}^{w}(\bX_{i})}+\widehat{Q}_{w}(\bX_{i},0)\Big]$},
\]
where $\widehat{Q}_{w}(\cdot,a)$ can be estimated by weighted least square for linear $Q$ functions with the estimated transfer weights, and $\hat{\pi}_{A}^{w}(\bX_{i})=\widehat{P}_{w}(A_{i}=1\mid\bX_{i})$ is the weighted propensity score which can be obtained by the weighted regression model with the estimated transfer weights using the experimental data. 

Let $\hat{\tau}_{i}^{w}$ be a generic weighted estimator of $\tau(\bX)$. Proposition \ref{prop:1} implies
an empirical risk function 
\begin{equation}
\begin{split}
 & \frac{1}{n}\sum_{i\in\rct}\hat{w}_{i}\mid\hat{\tau}_{i}^{w}\mid\{\indicator{(\hat{\tau}_{i}^{w}>0)}-d(\bX_{i};\be)\}^{2} \\ = & \frac{1}{n} \sum_{i\in\rct}\hat{w}_{i}\mid\hat{\tau}_{i}^{w}\mid l_{0-1}\left[\left\{ 2\indicator(\hat{\tau}_{i}^{w}>0)-1\right\} f(\bX_{i};\be)\right] \label{eq:empObj-w}
\end{split}
\end{equation}
to estimate $\eta$. 

It is challenging to optimize the objective function (\ref{eq:empObj-w})
since it involves a non-smooth non-convex objective function of $\be$.
One way is to optimize it directly using grid search or the genetic
algorithm in \citet{zhang2012robust} for learning linear decision
rules (implemented with \texttt{rgenoud} in R), but grid search becomes
untenable for a higher dimension of $\eta$, and the genetic algorithm
cannot guarantee a unique solution. Alternatively, one can replace
$l_{0-1}$ with a surrogate loss function, such as a hinge loss used
in the support vector machine \citep{Vapnik1998} and outcome weighted
learning \citep{zhao2012estimating}, or a ramp loss \citep{collobert2006trading}
which truncates the unbounded hinge loss by trading convexity for
robustness to outliers \citep{wu2007robust}. We adopt the smoothed
ramp loss function $l(u)$ proposed in \citet{zhou2017residual}, which retains
the robustness and also gains computational advantages due to being
smooth everywhere, defined in \eqref{eq:sramp loss}. $l(u)$ can be decomposed into the difference
of two smooth convex functions, $l(u)=l_{1}(u)-l_{0}(u)$, where $l_s(u)$ is defined in \eqref{eq:formula:decom}.

\begin{minipage}{0.49\textwidth}
  \begin{align}
l(u) & =\begin{cases}
0 & \text{if }u\geq1,\\
(1-u)^{2} & \text{if }0\leq u<1,\\
2-(1+u)^{2} & \text{if }-1\leq u<0,\\
2 & \text{if \ensuremath{u\leq-1}.}
\end{cases}\label{eq:sramp loss}
\end{align} 
\end{minipage}\hspace{0.5cm}
\begin{minipage}{0.38\textwidth}
\begin{align}
l_{s}(u) & =\begin{cases}
0 & \text{if }u\geq s,\\
(s-u)^{2} & \text{if }s-1\leq u<s,\\
2s-2u-1 & \text{if }u<s-1.
\end{cases}\label{eq:formula:decom}
\end{align}
\end{minipage}

Similarly to \citet{zhou2017residual}, we can apply the d.c. algorithm
\citep{le1997solving} to solve the non-convex minimization problem
by first decomposing the original objective function into the summation
of a convex component and a concave component and then minimizing
a sequence of convex subproblems. Applying (\ref{eq:formula:decom})
to (\ref{eq:empObj-w}), our empirical risk objective becomes
\begin{equation}
\underbrace{\frac{1}{n}\sum_{i\in\rct}\hat{w}_{i}\mid\hat{\tau}_{i}^{w}\mid l_{1}(u_{i})}_{\text{a convex function of \ensuremath{\be}}}+\underbrace{\frac{1}{n}\sum_{i\in\rct}\lf\{-\hat{w}_{i}\mid\hat{\tau}_{i}^{w}\mid l_{0}(u_{i})\ri\}}_{\text{a concave function of \ensuremath{\be} }},\label{formula:sramp obj}
\end{equation}
where $u_{i}=\lf\{2\indicator(\hat{\tau}_{i}^{w}>0)-1\ri\}f(\bX_{i};\be)$.
Let $C_{i}^{\text{cav}}=-\hat{w}_{i}\mid\hat{\tau}_{i}^{w}\mid l_{0}(u_{i})$
and 
\begin{equation}
\xi_{i}=\frac{\partial C_{i}^{\text{cav}}}{\partial u_{i}}=-\hat{w}_{i}\mid\hat{\tau}_{i}^{w}\mid\frac{\partial l_{0}(u_{i})}{\partial u_{i}},\label{formula:derivative}
\end{equation}
for $i\in\rct.$ Therefore, the convex subproblem at $(t+1)$th iteration
in the d.c. algorithm is to 
\begin{equation}
\underset{\be}{\text{min}}\;\;\;\frac{1}{n}\sum_{i\in\rct}\hat{w}_{i}\mid\hat{\tau}_{i}^{w}\mid l_{1}(u_{i})+\frac{1}{n}\sum_{i\in\rct}\xi_{i}^{(t)}u_{i},\label{formula:min gamma}
\end{equation}
which is a smooth unconstrained optimization problem and can be solved
with many efficient algorithms such as L-BFGS \citep{nocedal1980updating}.
Algorithm \ref{alg:dc} summarizes the proposed procedure to obtain
the minimizer $\hat{\eta}$. 
\begin{algorithm}[ht!]
\caption{The d.c. Algorithm for Finding Linear Decision Rule Parameters}
\label{alg:dc} \begin{algorithmic} \State Set a small value for
error tolerance, say $\epsilon=10^{-6}$; Initialize $\xi_{i}^{(0)}=2\mid\hat{\tau}_{i}\mid$.
\Repeat \State Obtain $\hat{\be}$ by solving (\ref{formula:min gamma})
with L-BFGS; \State Update $u_{i}=\lf\{2\indicator(\hat{\tau}_{i}>0)-1\ri\}f(\bX_{i};\hat{\be})$;
\State Update $\xi_{i}^{(t)}$ by (\ref{formula:derivative}). \Until
{$\mid\mid\mathbb{\mathbf{\bm{\xi}}}^{(t+1)}-\bm{\xi}^{(t)}\mid\mid_{\infty}\leq\epsilon$}
\State \textbf{Return} $\hat{\be}$. \end{algorithmic}
\end{algorithm}


\subsection{Cross-validation\label{sec:cv} }

We have discussed various approaches for the estimation of transfer
weights and individual contrast function. Moreover, although transfer
weights can correct the selection bias of the experimental data, they
also increase the variance of the estimator compared to the unweighted
counterpart. The effect of weighting on the precision of estimators
can be quantified by the effective sample size originated in survey
sampling \citep{kish1992weighting}. If the true contrast function
can be approximated well by linear models, the estimated ITR with
transfer weights may not outperform the one without weighting. As
a result, for a given application, it is critical to develop a data-adaptive
procedure to choose the best method among the aforementioned weighted
estimators and whether or not using transfer weights. 

Toward this end, we propose a cross-validation procedure. To proceed,
we split the sample into two parts, one for learning the ITRs and
the other for evaluating the estimated ITRs. We also adopt multiple
sample splits in order to increase the robustness since single split
might have dramatic varying results due to randomness. For evaluation
of an ITR $d_{\be}$, we estimate the value for the given ITR by the
weighted AIPW estimator in (\ref{eq:w_aipw}) with the estimated nonparametric
transfer weights $\hat{w}$,
\begin{equation}
\begin{split}
\widehat{V}(d_{\be};\hat{w},\rct)=\sum_{i\in\rct}\hat{w}_{i} \Bigg(& \Big[\frac{A_{i}d(\bX_{i};\be)}{\hat{\pi}_{A}^{w}(\bX_{i})}+\frac{(1-A_{i})\left\{ 1-d(\bX_{i};\be)\right\} }{1-\hat{\pi}_{A}^{w}(\bX_{i})}\Big]\left[Y_{i} -\widehat{Q}_{w}\left\{ \bX_{i},d(\bX_{i};\be)\right\} \right]\\ &-\widehat{Q}_{w}\left\{ \bX_{i},d(\bX_{i};\be)\right\}\Bigg).\label{eq:w_aipw}
\end{split}
\end{equation}

Since we are more interested in the comparisons among different weighting
and unweighting methods instead of different contrast function estimation
methods, thus we fix an estimation method for the contrast function
at first, and then use cross-validation to choose among the set of
candidate methods $\mathcal{G}$. This $\mathcal{G}$ contains the
proposed learning method with nonparametric transfer weights described
in Section \ref{subsec:Nonparametric-approach}, and unweighted estimation
method. If the target population size $N$ is available, we can also
include the weighted estimators with the two parametric approaches
introduced in Section \ref{subsec:Parametric-approach} into $\mathcal{G}$.
Then we can get the value estimates averaged over all the sample splits
for each learning method in $\mathcal{G}$, and return the method
with the highest value estimates. The more explicit description of
the procedure is presented in Algorithm \ref{alg2}.

\begin{algorithm}[h!]
\caption{Cross-validation with Multi-Sample Splits}
\label{alg2} \begin{algorithmic} \State \textbf{Input} The experimental
data $\mathcal{D}_{n}=\{(\bX_{i},A_{i},Y_{i})\}_{i\in\mathcal{I}_{n}}$,
the RWD $\mathcal{D}_{m}=\{\bX_{i}\}_{i\in\mathcal{I}_{m}}$, the
number of sample splits $B$, the set of candidate methods need to
compare $\mathcal{G}$, the target sample size $N$ is optional (needed
if $\mathbb{\mathcal{G}}$ contains parametric weighting methods).

\For{b = 1, \dots , $B$} \State Randomly split $\mathcal{D}_{n}$
into two equal-size parts as the experimental training and testing
data: $\mathcal{D}_{n}^{\text{train}}$ and $\mathcal{D}_{n}^{\text{test}}$;
let $\mathcal{I}_{n}^{\text{test}}$ be the index set of $\mathcal{D}_{n}^{\text{test}}$.
Similarly, randomly split $\mathcal{D}_{m}$ into two equal-size parts
$\mathcal{D}_{m}^{\text{train}}$ and $\mathcal{D}_{m}^{\text{test}}$;

\State Estimate ITRs with all methods in $\mathcal{G}$ based on
$\mathcal{D}_{n}^{\text{train}}$ and $\mathcal{D}_{m}^{\text{train}}$
(use $N/2$ as the target population size if $\mathbb{\mathcal{G}}$
contains parametric weighting methods). Denote the estimated ITRs
as $\{\hat{d}_{g}^{\text{train}}\}_{g\in\mathcal{G}}$;

\State Obtain the nonparametric transfer weights $\{\widetilde{w}_{i}\}_{i\in\mathcal{I}_{n}^{\text{test}}}$
based on the test data $\mathcal{D}_{n}^{\text{test}}$ and $\mathcal{D}_{m}^{\text{test}}$;

\State Estimate the value of the estimated ITRs with weighted AIPW
$\widehat{V}(\hat{d}_{g}^{\text{train}};\widetilde{w},\mathcal{I}_{n}^{\text{test}})$ 
, denoted as as $\widehat{V}_{g}^{(b)}, g\in\mathcal{G}$.

\EndFor \State Average the estimated values over $B$ splits for
each estimated ITR, denoted as $\overline{V}_{g}=B^{-1}\sum_{b=1}^{B}\widehat{V}_{g}^{(b)}$,
$g\in\mathcal{G}$. \State \textbf{Return} $\underset{g\in\mathcal{G}}{\arg\max}=\overline{V}_{g}$.
\end{algorithmic}
\end{algorithm}

\section{Asymptotic Properties \label{sec:asy} }

In this section, we provide a theoretical guarantee that the estimated
ITR within a certain class $\mathcal{F}$ is consistent for the true
population optimal ITR within $\mathcal{F}$. Toward this end, we
show that the expected risk of the estimated ITR is consistent for
that of the true optimal ITR within $\mathcal{F}$. In particular,
we consider the ITRs learned by the nonparametric transfer weights,
weighted AIPW estimator $\hat{\tau}_{\text{aw}}^{w}$ with the logistic
regression model $\pi_{A}(\bX;\gamma)$ and the linear $Q$ function.
The extensions to other cases with parametric transfer weights and
regression and IPW estimators are straightforward. 

We introduce more notation. Let $\theta=(\mathbf{\gamma^{\transpose},}\beta^{\transpose})^{\transpose}$
and let the contrast function
\begin{equation}
\resizebox{0.92\hsize}{!}{$
\tau_{\text{aw}}^{\theta}(\bX)=\frac{\indicator(A=1)\left\{ Y-Q(\bX,1;\beta)\right\} }{\pi_{A}(\bX;\gamma)}+Q(\bX,1;\beta)-\frac{\indicator(A=0)\{Y-Q(\bX,0;\beta)\}}{1-\pi_{A}(\bX;\gamma)}-Q(\bX,0;\beta)$}.\label{eq:tau_theta}
\end{equation}
Denote the general smooth ramp loss function $l^{\zeta}(u)=l(\zeta_{1}u)/\zeta_{2}$,
where $\zeta_{1},\zeta_{2}>0$, $l$ is defined in \eqref{eq:sramp loss},
so it is easy to see that when $\zeta=(\zeta_{1},\zeta_{2})\rightarrow\zeta^{*}\define(+\infty,2)$,
the general smooth ramp loss function will converge to the 0-1 loss
function, i.e., $l^{\zeta}(u)\rightarrow l_{0-1}(u),\forall u$. Note
that the general smooth ramp loss function can also be written as
the difference of two convex functions, $l^{\zeta}(u)=l_{1}(\zeta_{1}u)/\zeta_{2}-l_{0}(\zeta_{1}u)/\zeta_{2}$,
thus the d.c. algorithm in Algorithm \ref{alg:dc} can be easily generalized
to $l^{\zeta}(u)$. Recalling the loss function $\mathcal{\mathcal{L}}(\tau;\be,l_{0-1})$
and the corresponding $\mathcal{F}$-risk $\mathcal{R}_{\mathcal{F}}(\tau;\be,l_{0-1})$
defined in \eqref{eq:obj_0_1_loss}, we replace\textbf{ $\tau$} with
$\tau_{\text{aw}}^{\theta}$ and obtain:
\[
\resizebox{0.98\hsize}{!}{$
\mathcal{R}_{\mathcal{F}}(\tau_{\text{aw}}^{\theta};\be,l_{0-1})=\pr\left\{ \mathcal{\mathcal{L}}\left(\tau_{\text{aw}}^{\theta};\be,l_{0-1}\right)\right\} ,\:\mathcal{\mathcal{L}}\left(\tau_{\text{aw}}^{\theta};\be,l_{0-1}\right)=\mid\tau_{\text{aw}}^{\theta}(\bX)\mid l_{0-1}\left([2\indicator\{\tau_{\text{aw}}^{\theta}(\bX)>0\}-1]f(\bX;\be)\right),\:\forall\be\in\mathcal{F}$},
\]
denoted as $\mathcal{R}_{\mathcal{F}}(\be)$ and $\mathcal{L}(\be,\theta)$
respectively for simplicity. We define the minimal $\mathcal{F}$-risk
as $\mathcal{R}_{\mathcal{F}}^{*}=\inf_{\be\in\mathcal{F}}\mathcal{R}_{\mathcal{F}}(\be)$.
Similarly, replacing $l_{0-1}$ in $\mathcal{R}_{\mathcal{F}}(\be)$
and $\mathcal{L}(\be,\theta)$ with the general smooth ramp loss $l^{\zeta}$,
we can obtain $\mathcal{R}_{\mathcal{F}}(\tau_{\text{aw}}^{\theta};\be,l^{\zeta})$
and $\mathcal{\mathcal{L}}\left(\tau_{\text{aw}}^{\theta};\be,l^{\zeta}\right)$,
denoted as as $\mathcal{R}_{\zeta\mathcal{F}}(\be)$ and $\Leps(\be,\theta)$,
respectively. Likewise, the minimal $\mathcal{F}$-risk with $l^{\zeta}$
is defined as $\mathcal{R}_{\zeta\mathcal{F}}^{*}=\inf_{\be\in\mathcal{F}}\mathcal{R}_{\zeta\mathcal{F}}(\be)$. 

The goal is to show that the decision rule is $\mathcal{F}$-consistency
\citep{bartlett2006convexity}, i.e., 
\[
\lim_{N\rightarrow\infty}\mathcal{R}_{\mathcal{F}}(\hat{\be})=\mathcal{R}_{\mathcal{F}}^{*},
\]
where $\hat{\be}=\arg\min_{\be\in\mathcal{F}}\sum_{i=1}^{N}S_{i}\hat{w}_{i}\Leps(\be,\hat{\theta}_{w})$,
where $\hat{\theta}_{w}=(\hat{\gamma}_{w}^{\transpose},\hat{\beta}_{w}^{\transpose})^{\transpose}$, the estimated logistic regression coefficients $\hat{\gamma}_{w}=\arg\max_{\gamma}\sum_{i=1}^{N}\hat{w}_{i}S_{i}[A_{i}\log\pi_{A}(\bX_{i};\gamma)+(1-A_{i})\log\left\{ 1-\pi_{A}(\bX_{i};\gamma)\right\}]$
which is converge in probability to $\arg\max_{\gamma}\expect[A\log\pi_{A}(\bX;\gamma)+(1-A_{i})\log\left\{ 1-\pi_{A}(\bX;\gamma)\right\}]$ $\define\gamma^{*}$
as $N\rightarrow\infty$ under regular conditions, and the estimated
outcome regression coefficients $\hat{\beta}_{w}=\arg\min_{\beta}\sum_{i=1}^{N}\hat{w}_{i}S_{i}(Y_{i}-\beta^{\transpose}L_{i})^{2}$,
where $L_{i}=(1,\bX_{i}^{\transpose},A_{i},A_{i}\bX_{i}^{\transpose})$,
thus we have 
\[
\begin{split}\hat{\beta}_{w} & =\left(\sum_{i=1}^{N}N\hat{w}_{i}S_{i}L_{i}L_{i}^{\transpose}\right)^{-1}\sum_{i=1}^{N}N\hat{w}_{i}S_{i}L_{i}Y_{i}\rightarrow\{\expect(LL^{\transpose})\}^{-1}\expect(LY)\define\beta^{*},\end{split}
\]
in probability as $N\rightarrow\infty$. The convergences of $\hat{\gamma}_{w}$
and $\hat{\beta}_{w}$ are implied from the weak law of large numbers
and the result $\max_{i}\mid N\hat{w}_{i}-1/\pi_{S}(\bX_{i})\mid=o_{p}(1)$
proved in Theorem 2 in \citet{wang2020minimal}. Let $\theta^{*}=(\gamma^{*\transpose},\beta^{*}{}^{\transpose})^{\transpose}$,
thus we have 
\begin{equation}
\thetaw\xrightarrow{p}\theta^{*}.\label{eq: thetahat}
\end{equation}
It is easy to see that $\mathcal{R}_{\zeta\mathcal{F}}(\be)$ will
converge to $\mathcal{R}_{\mathcal{F}}(\be)$ when $\zeta\rightarrow\zeta^{*},$
therefore, to obtain $\mathcal{F}$-consistency, it is suffice to
show that $\lim_{N\rightarrow\infty}\mathcal{R}_{\zeta\mathcal{F}}(\hat{\be})=\mathcal{R}_{\zeta\mathcal{F}}^{*}$
which is stated in the theorem below. 

\begin{theorem} \label{thm1} Under Assumptions S1--S10, 
we have $\lim_{N\rightarrow\infty}\mathcal{R}_{\zeta\mathcal{F}}(\hat{\be})=\mathcal{R}_{\zeta\mathcal{F}}^{*}.$
\end{theorem} 

\section{Simulation study\label{sec:sim} }

We evaluate the finite-sample performances of the proposed estimators
with a set of simulation studies. We first generate a target population
of a size $N=10^{6}$. The covariates are generated by $\bX_{i}=(X_{i1},X_{i2})^{\intercal}\sim N(\mathbf{1}_{2\times1},I_{2\times2})$,
and the potential outcome is generated by $Y_{i}^{*}(a)\mid\bX_{i}=1+2X_{i1}+3X_{i2}+a\tau(\bX_{i})+\epsilon_{i}(a),$
where $\epsilon_{i}(a)$ i.i.d. follows $N(0,0.5^{2})$ for $a=0,1.$
We consider three contrast functions: (I). $\tau(\bX_{i})=\arctan\{\exp(1+X_{i1})-3X_{i2}-5\};$ (II). $\tau(\bX_{i})=\cos(1)+\cos(X_{i1})+\cos(X_{i2})-3/2;$
(III). $\tau(\bX_{i}) = 1+2X_{i1}+3X_{i2}$.

In order to evaluate the necessity and performances of weighting methods, we consider two categories of data generating models. One is the scenarios where the true optimal policies are not in linear forms , i.e., the settings (I) and (II), while the other one is considering the true optimal policy is in a linear form as the setting (III). We start with the first two scenarios. From the data visualization, we observed that the covariates shifted
between the experimental data and RWD to a varying extent in the two
generative models: the covariates generated by (I) have similar distributions
between the experimental data and RWD, while those generated by (II)
have very different ones. Then we generate the RWD by randomly sample
$m=5000$ subjects from the target population $\{\bX_{i}\}_{i=1}^{N}$,
denoted as $\{\bX_{i}\}_{i\in\mathcal{I}_{m}}$. To form the experimental
sample, we generate the indicator of selection according to $S_{i}\mid\bX_{i}\sim\text{Bernoulli}\left[\exp(\alpha_{0}+\alpha_{1}^{^{\intercal}}\bX_{i})/\{1+\exp(\alpha_{0}+\alpha_{1}^{\intercal}\bX_{i})\}\right]$,
where $\alpha_{1}=(1,-2)^{\intercal}$ and $\alpha_{0}=-8$ which gives the average sample sizes
of the experimental data over the simulation replicates round 1386, respectively. In the experimental data denoted
as $\rct$, the treatments are randomly assigned according to $A_{i}\sim\mathrm{Bernoulli}(0.5)$,
and then the actual observed outcomes $Y_{i}=Y_{i}^{*}(A_{i})$, $i\in\mathcal{I}_{n}$. 

For each setting, we compare the following estimators:

1. \emph{``$w_{1}$'': }our proposed weighted estimator with parametric
weights learned by the modified maximum likelihood introduced in Section
\ref{subsec:Parametric-approach}.

2. \emph{``$w_{2}$'': }our proposed weighted estimator with parametric
weights learned by the estimating equations introduced in Section
\ref{subsec:Parametric-approach}.

3. \emph{``cv''}: our proposed weighted estimator using cross-validation
procedure introduced in Section \ref{sec:cv}, and we set the number
of sample splits as 10 in all the simulation studies.

4. \emph{``np''}: our proposed weighted estimator with nonparametric
weights.

5. \emph{``unweight'': }only using the experimental data to learn
the optimal linear ITRs.

6.\emph{ ``bm''}: using the weighted estimator but we replace
the estimated transfer weights $\hat{w}_{i}$ with the normalized
true weights $w_{i}=\{\pi_{S}(\bX_{i})\}^{-1}/\sum_{i\in\rct}\{\pi_{S}(\bX_{i})\}^{-1}$,
and replace the estimated contract functions $\hat{\tau}_{i}$ with the true contrast values $Y_i^*(1)-Y_i^*(0)$.

7. \emph{``Imai''}: the method proposed in \citet{imai2013estimating}.
They construct the transfer weights by fitting a Bayesian additive
regression trees (BART) model using covariates as predictors and whether
to be in the experimental data as outcomes, and then use a variant
of SVM to estimate the conditional treatment effect (implemented in the
R package \texttt{FindIt}). Then given $\bX$, we can assign treatment
$1$ if its estimated conditional treatment effect is positive, and
$0$ otherwise. 

8. \emph{``DBN''}: the method proposed in \citet{mo2020learning}, where they propose to use distributionally robust ITRs, that is, value functions are evaluated under all testing covariate distributions that are "close" to the training distribution, and the worse-case takes a minimal one, and then the ITRs are learned by maximizing the defined worst-case value function.

For the estimators 1. -- 5. in the above, we estimate the contrast function using $\tau_\text{aw}$ with linear $Q$ functions and constant propensity scores estimated via the proportion
of $A=1$ in the experimental data, $n^{-1}\sum_{i\in\rct}A_{i}$.

To study the impact of model misspecification of $\pi_{S}(\bX)$,
we consider to estimate $\alpha$ by fitting logistic regression wrongly:
$\pi_{S}(\bX;\alpha)=\exp(\alpha_{0}+\alpha_{1}^{\intercal}\widetilde{\bX})/\{1+\exp(\alpha_{0}+\alpha_{1}^{\intercal}\widetilde{\bX})\}$,
where $\widetilde{\bX}=(X_{1}^{2},X_{2}^{2})$. And as described in the proposed methods, only estimators 1. -- 3. need to specify the sampling model. To measure the performances of the different approaches, we use the mean value MSE defined as $N^{-1}\sum_{i=1}^N \left(Y_i^*\{d_{\eta}(\bX_i)\} - Y_i^*[\indicator\{\tau(\bX_i)>0\}]\right)^2$. Thus the lower the MSE, the better performances the approach has.

\begin{figure}[ht!]
\includegraphics[width=1\textwidth]{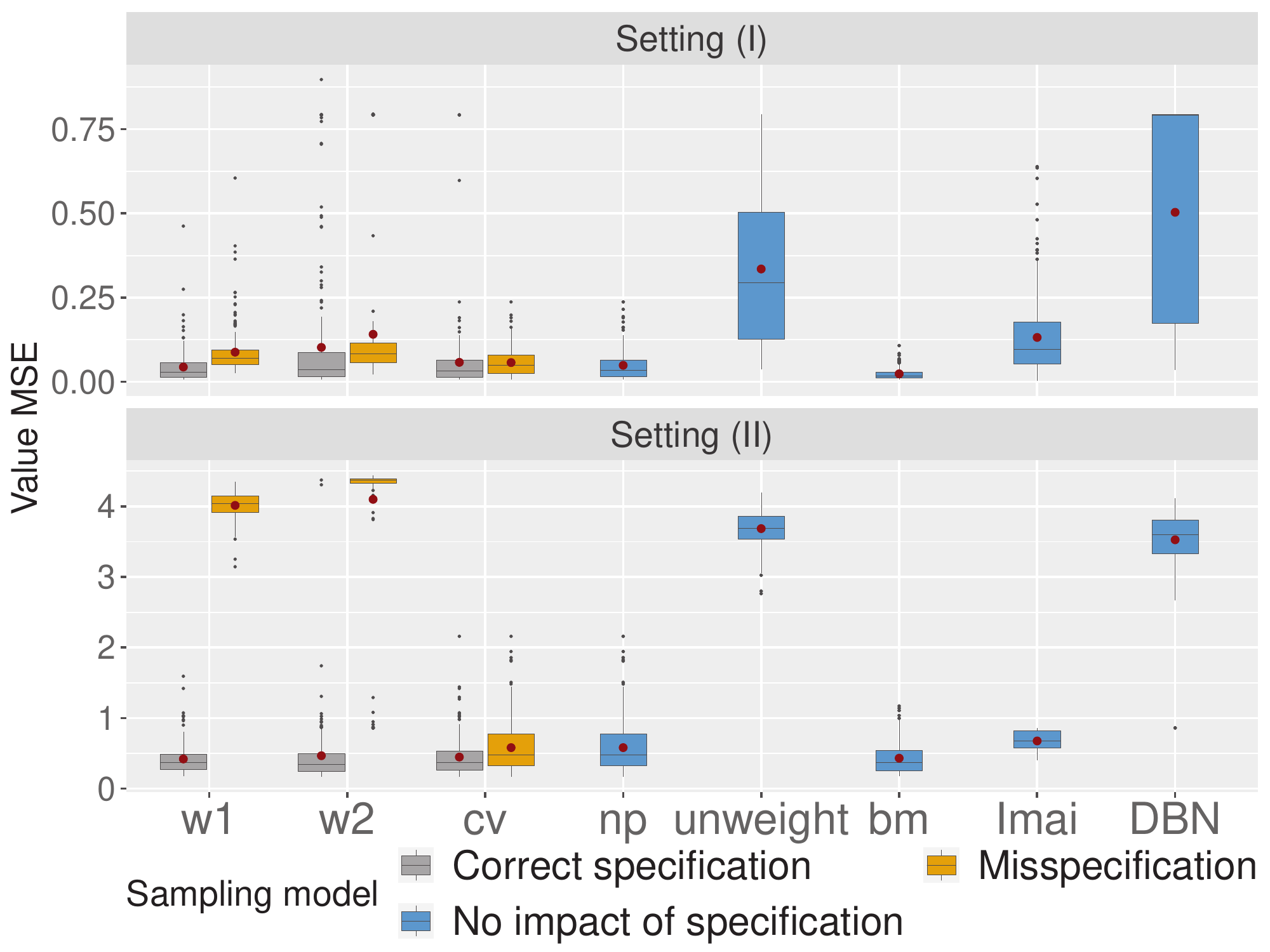}
\caption{Value MSE boxplots of different methods for the generative model (I) and (II). The red point in
each box is the average of the value MSE.}
\label{fig:similar & diff}
\end{figure}

The results are summarized in Figure \ref{fig:similar & diff},
the dark red point in each boxplot is the mean value MSE averaged
over 200 Monte Carlo replicates.
For the generative model (I) shown in Figure \ref{fig:similar & diff}, unweighted and DBN rules perform worst in all the cases. The second parametric weighting method is relatively robust compare with the first parametric one when the sampling model is misspecified; this is reasonable since the first one uses maximum likelihood estimation which is more reliable on the correctness of the parametric model specification. When the training data and testing data are similarly distributed, the performances of parametric, nonparametric weighting method and the cross-validation procedure are similar; in this case, sampling model misspecification does not have much influence on the rule learning.

For the generative model (II) shown in Figure \ref{fig:similar & diff}, among the performances
of weighted and unweighted estimators, we can see the weighted estimators
outperform the unweighted estimators when model correctly specified;
the nonparametric weighted estimator outperforms other estimators
and the performance of the parametric weighted estimators have dramatically
decreased when the sampling model misspecified. These findings make sense, because when the more different the training and testing data are, the more crucial the bias correction is. Thus sampling model misspecification might have huge influence on the learned rules. 

From the both generative models, we can see that the nonparametric weighting and cross-validation procedure are more robust when the sampling model is unknown and recommended in practice use. As for DBN estimator, it performs badly in terms of MSE magnitude or variability, one possible reason might be that the tuning parameters when measuring the "closeness" of two distributions need to be chosen more carefully in practice \cite{mo2020learning}. As for the comparison with Imai also has relative good performance among all estimators and has relative small variability.

\begin{figure}[ht!]
\includegraphics[width=0.9\textwidth]{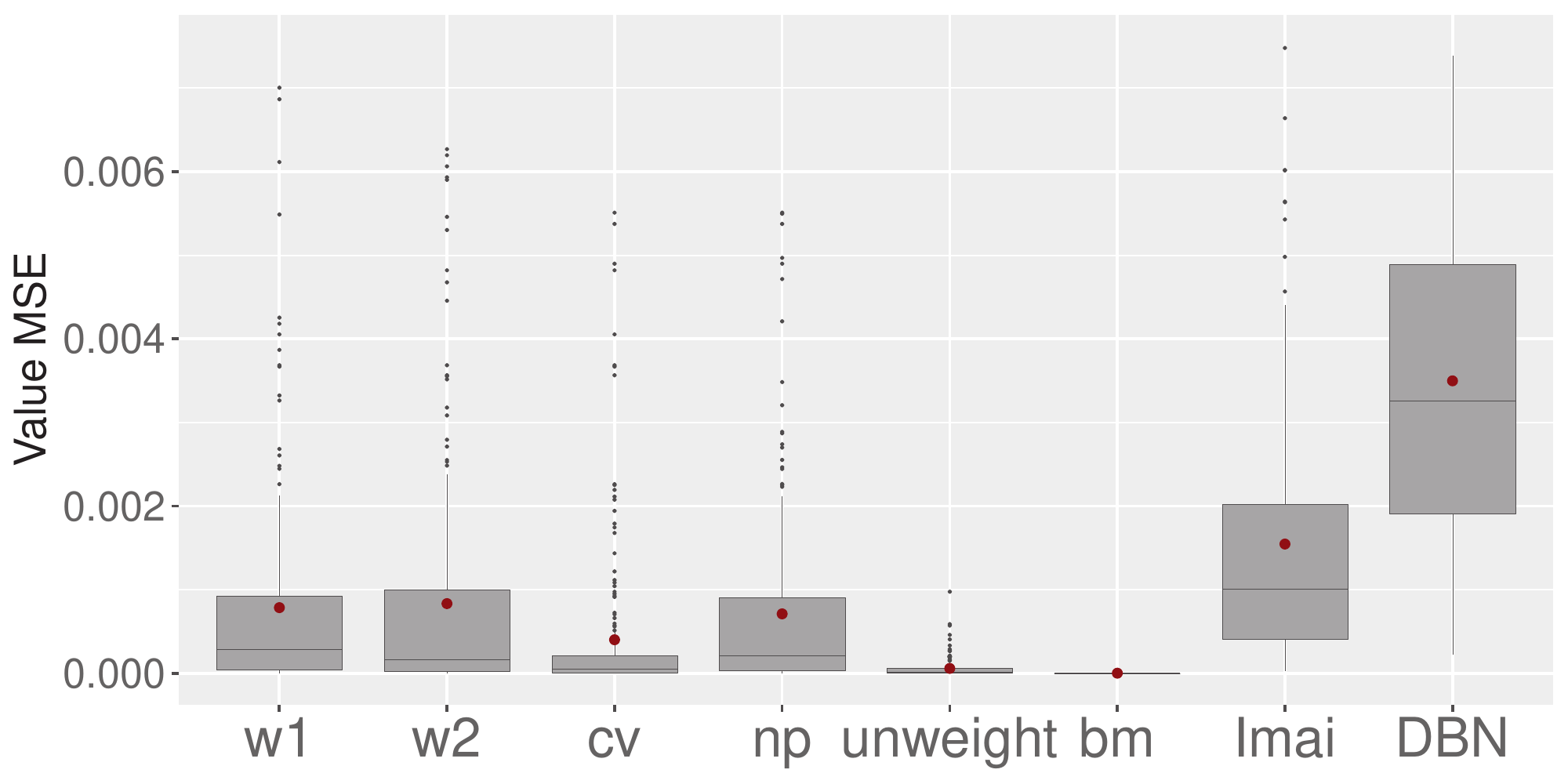}
\caption{Value MSE boxplots of different methods for the generative model (III).}
\label{fig:linear}
\end{figure}

As for setting (III), we consider the correctly specified sample score model. As shown in Figure \ref{fig:linear}, unweighted method performances the best, close to the benchmark. This 
demonstrates that when the true optimal polices are in linear forms, then learning with only the experimental data performances the best. Thus weighting is not a necessity, and it even will induce the variance of the estimators. The other competitive estimator is using cross-validation. Therefore, when we have no idea of the true data generating models, using CV is a good choice in practice. 

\section{Real Data Application\label{sec:application} }

We apply the proposed methods to personalized recommendations of a
job training program on post-market earning. Our analysis is based
on two data sources (downloaded from \url{https://users.nber.org/~rdehejia/data/.nswdata2.html}): an experimental dataset from the National Supported
Work (NSW) program \citep{lalonde1986evaluating} and a non-experimental
dataset from the Current Population Survey (CPS; \citet{lalonde1986evaluating}).
The NSW dataset includes $185$ individuals who received the job training
program ($A=1$) and $260$ individuals who did not receive the job
training program ($A=0$), and the CPS dataset includes $15992$ individuals
without the job training program. Let $\bX$ be a vector including
an intercept $1$ and the pre-intervention covariates: age, education,
Black (1 if black, 0 otherwise), Hispanic (1 if Hispanic, 0 otherwise),
married (1 if married, 0 otherwise), whether having high school degree
(1 if no degree, 0 otherwise), earning in 1974, and earning in 1975.
The outcome variable is the earning in 1978. We transfer the earnings
in the year 1974, 1975, and 1978 by taking the logarithm of the earnings
plus one, denoted as ``log.Re74'', ``log.Re75'' and ``log.Re78''.
We aim to learn a linear rule $d(\bX;\be)=\indicator(\bX^{\transpose}\eta>0)$
tailoring the recommendation of the job training program to an individual
with characteristics $\bX$. 

The CPS sample is generally representative of the target population;
while the participants to the NSW program may not represent the general
population. Table \ref{tab: real data} contrasts sample averages
of each pre-intervention covariate in the NSW and CPS samples and
shows that the covariates are highly imbalanced between the two samples.
On average, the NSW sample is younger, less educated, more black people,
fewer married, more without high school degree, fewer earnings in
1974 and 1975.\textbf{ }Because of imbalance, ITRs cannot be learned
based only on the NSW sample, thus we are motivated to use the proposed
weighted transfer learning approaches to learn optimal linear ITRs
which benefit the target population most.

\begin{table}[ht]
\caption{Sample averages of pre-intervention variables in the NSW and CPS samples
and results from the weighted transfer and unweighted learning methods:
The first two rows are sample averages of each pre-intervention covariate
in the NSW and CPS samples, respectively. The last two rows are estimated
linear coefficients of each corresponding variable in the linear decision
rule from the weighted transfer and unweighted learning methods, respectively.}
\label{tab: real data}

\centering 

\resizebox{\textwidth}{!}{
\begin{tabular}{cccccccccc}
\hline 
 & intercept & Age & Educ & Black & Hisp & Married & Nodegr & log.Re74 & log.Re75\tabularnewline
\hline 
experiment NSW & 1 & 25.37 & 10.20 & 0.83 & 0.09 & 0.17 & 0.78 & 2.27 & 2.70\tabularnewline
Non-experiment CPS & 1 & 33.23 & 12.03 & 0.07 & 0.07 & 0.71 & 0.30 & 8.23 & 8.29\tabularnewline
\hline 
 & $\eta_{0}$ & $\eta_{1}$ & $\eta_{2}$ & $\eta_{3}$ & $\eta_{4}$ & $\eta_{5}$ & $\eta_{6}$ & $\eta_{7}$ & $\eta_{8}$\tabularnewline
\hline 
Unweighted learning $\hat{\eta}_{{\rm u}}$ & -1 & 0.03 & 0.07 & 0.81 & 2.31 & 1.89 & -0.62 & -0.60 & 0.47\tabularnewline
Weighted transfer learning $\hat{\eta}_{{\rm w}}$ & -1 & 0.02 & 0.05 & 0.66 & 0.36 & 0.14 & 0.28 & -0.27 & 0.15\tabularnewline
\hline 
\end{tabular}}
\end{table}

We consider the unweighted and weighted transfer learning methods.
For the weighted transfer learning method, we consider the nonparametric
weighting approach, because the parametric approach requires the target
population size $N$ which is unknown in this application. To estimate
the weighted and unweighted decision rules, we use the AIPW estimator
$\hat{\tau}_{\text{aw}}$ with linear Q functions since it outperforms
other competitors in the simulation study in Section \ref{sec:sim}.
We also implement the cross-validation procedure to choose between
the unweighted and weighted transfer learning methods. 

Table \ref{tab: real data} presents the estimated linear coefficients
of each corresponding variable in the linear decision rule from the
weighted transfer and unweighted learning methods, respectively. Moreover,
from the cross-validation procedure, the weighted transfer learning
method produces a smaller population risk than the unweighted learning
method and therefore is chosen as the final result. Compared with
the learned decision rule with only the experiment data NSW which
oversamples the low-income individuals, individuals without high school
degrees (i.e. Nodegr = 1) are more likely to have higher outcomes
if offered the job training program in the target population. 

To assess the performances of the learned rules for the target population,
following \citet{kallus2017recursive}, we create a test sample that
is representative of the target population. We sample $100$ individuals
at random from the CPS sample to form a control group and then find
their closest matches in the NSW treated sample to form the treatment
group (use \texttt{pairmatch} in R package \texttt{optmatch}). The
test sample thus contains the $100$ matched pairs, mimicking a experimental
sample where binary actions are randomly assigned with equal probabilities.
We use random forest on the test data to estimate $Q$ function $\expect(Y\mid\bX,a)$,
denoted the estimator as $\widehat{Q}_{\text{test}}(\cdot,a),$ $a=0,1$,
then we can use $\sum_{i\in\text{CPS}}\widehat{Q}_{\text{test}}\left\{ \bX_{i},d(\bX_{i};\be)\right\} /\mid\text{CPS}\mid$
to estimate the value of an ITR $d(\cdot;\be)$, where $\mid\text{CPS}\mid$
is the sample size of CPS. Besides the weighted linear rule $d(\cdot;\hat{\be}_{w})$
and the unweighted linear rule $d(\cdot;\hat{\be}_{u})$ learned from
the above, we also compare the Imai method introduced in \citet{imai2013estimating}:
first to estimate the conditional probability of being selected into
NSW given $\bX$ from the total sample which consists of NSW and CSP
using BART model, and then to estimate the conditional treatment effect
weighted by the inverse of the sampling probability. Then the Imai
rule will recommend the job training program if the conditional treatment effect is positive and will not recommend it otherwise. We replicate this sampling test data procedure 100 times, estimate values of the three ITRs for each replicate, and get the averaged estimated values (standard error) as follows: 8.395 (0.009), 6.106 (0.011), and 8.049 (0.006) for $d(\cdot;\hat{\be}_{w})$, $d(\cdot;\hat{\be}_{u})$,
and Imai respectively, thus the transfer weighted rule has the highest
value estimate. Although this evaluation method needs assumption of
no unmeasured confounders for the non-experimental data CPS, which
is not needed in our learning procedure, it can still bring us confidence, to some degree, that the weighted linear rule has better generalizability
to the target population.

\section{Discussion\label{sec:Discussion}}

We develop a general framework to learn the optimal interpretable
ITRs for the target population by leveraging no unmeasured confounding
in the experimental data and the representativeness of covariates
in the RWD. The proposed procedure is easy to implement. By constructing
transfer weights to correct the distribution of covariates in the
experimental data, we can learn the optimal interpretable ITRs for
the target population. Besides, we provide a data-adaptive procedure
to choose among different weighting methods by modified cross-validation.
Moreover, we establish the theoretical guarantee for the consistency
of the proposed estimator of ITRs. 

Some future directions worth considering. One is to quantify the uncertainty
of the estimated ITRs and therefore to conduct statistical inference.
Moreover, in some scenarios, beside the covariate information, comparable
treatment and outcome information are also available in the RWD. Unlike
the randomized experiment, the treatment assignment in the RWD is
determined by the preference of physicians and patients. In this case,
an interesting topic is to develop integrative methods that can leverage
the strengths of both data sources for efficient estimation of the
ITRs. 

\bibliographystyle{dcu}

\bibliography{ref}

@article{hainmueller2012entropy,
  title={Entropy balancing for causal effects: A multivariate reweighting method to produce balanced samples in observational studies},
  author={Hainmueller, Jens},
  journal={Political analysis},
  pages={25--46},
  year={2012},
  publisher={JSTOR}
}

@article{zhao2019robustifying,
  title={Robustifying trial-derived optimal treatment rules for a target population},
  author={Zhao, Ying-Qi and Zeng, Donglin and Tangen, Catherine M and LeBlanc, Michael L},
  journal={Electronic journal of statistics},
  volume={13},
  number={1},
  pages={1717},
  year={2019},
  publisher={NIH Public Access}
}

@article{mo2020learning,
  title={Learning optimal distributionally robust individualized treatment rules},
  author={Mo, Weibin and Qi, Zhengling and Liu, Yufeng},
  journal={Journal of the American Statistical Association},
  pages={1--16},
  year={2020},
  publisher={Taylor \& Francis}
}

@article{imai2013estimating,
  title={Estimating treatment effect heterogeneity in randomized program evaluation},
  author={Imai, Kosuke and Ratkovic, Marc and others},
  journal={The Annals of Applied Statistics},
  volume={7},
  number={1},
  pages={443--470},
  year={2013},
  publisher={Institute of Mathematical Statistics}
}

@Article{optmatch,
  author = {Ben B. Hansen and Stephanie Olsen Klopfer},
  title = {Optimal full matching and related designs via network flows},
  journal = {Journal of Computational and Graphical Statistics},
  volume = {15},
  number = {3},
  pages = {609--627},
  year = {2006},
}

@Manual{findit,
  title = {FindIt: Finding Heterogeneous Treatment Effects},
  author = {Naoki Egami and Marc Ratkovic and Kosuke Imai},
  year = {2019},
  note = {R package version 1.2.0},
  url = {https://CRAN.R-project.org/package=FindIt},
}

@article{zhou2017causal,
	title={Causal nearest neighbor rules for optimal treatment regimes},
	author={Zhou, Xin and Kosorok, Michael R},
	journal={arXiv preprint arXiv:1711.08451},
	year={2017}
}

@article{greenland1990randomization,
	title={Randomization, statistics, and causal inference},
	author={Greenland, Sander},
	journal={Epidemiology},
	pages={421--429},
	year={1990},
	publisher={JSTOR}
}

@article{moodie2014q,
	title={Q-learning: Flexible learning about useful utilities},
	author={Moodie, Erica EM and Dean, Nema and Sun, Yue Ru},
	journal={Statistics in Biosciences},
	volume={6},
	number={2},
	pages={223--243},
	year={2014},
	publisher={Springer}
}

@article{qian2011performance,
	title={Performance guarantees for individualized treatment rules},
	author={Qian, Min and Murphy, Susan A},
	journal={Annals of statistics},
	volume={39},
	number={2},
	pages={1180},
	year={2011},
	publisher={NIH Public Access}
}

@article{kang2014combining,
	title={Combining biomarkers to optimize patient treatment recommendations},
	author={Kang, Chaeryon and Janes, Holly and Huang, Ying},
	journal={Biometrics},
	volume={70},
	number={3},
	pages={695--707},
	year={2014},
	publisher={Wiley Online Library}
}

@article{lalonde1986evaluating,
	title={Evaluating the econometric evaluations of training programs with experimental data},
	author={LaLonde, Robert J},
	journal={The American economic review},
	pages={604--620},
	year={1986},
	publisher={JSTOR}
}

@article{kish1992weighting,
	title={Weighting for unequal Pi.},
	author={Kish, Leslie and Stat, J Official},
	year={1992}
}

@article{kouw2018introduction,
	title={An introduction to domain adaptation and transfer learning},
	author={Kouw, Wouter M and Loog, Marco},
	journal={arXiv preprint arXiv:1812.11806},
	year={2018}
}

@article{murphy2005generalization,
	title={A generalization error for Q-learning},
	author={Murphy, Susan A},
	journal={Journal of Machine Learning Research},
	volume={6},
	number={Jul},
	pages={1073--1097},
	year={2005}
}

@article{horvitz1952generalization,
	title={A generalization of sampling without replacement from a finite universe},
	author={Horvitz, Daniel G and Thompson, Donovan J},
	journal={Journal of the American statistical Association},
	volume={47},
	number={260},
	pages={663--685},
	year={1952},
	publisher={Taylor \& Francis Group}
}

@book{chakraborty2013statistical,
	title={Statistical methods for dynamic treatment regimes},
	author={Chakraborty, Bibhas},
	year={2013},
	publisher={Springer}
}

@article{cole2010generalizing,
  title={Generalizing evidence from randomized clinical trials to target populations: the ACTG 320 trial},
  author={Cole, Stephen R and Stuart, Elizabeth A},
  journal={American journal of epidemiology},
  volume={172},
  number={1},
  pages={107--115},
  year={2010},
  publisher={Oxford University Press}
}

@article{rothwell2005external,
  title={External validity of randomised controlled trials:“to whom do the results of this trial apply?”},
  author={Rothwell, Peter M},
  journal={The Lancet},
  volume={365},
  number={9453},
  pages={82--93},
  year={2005},
  publisher={Elsevier}
}

@article{hartman2015sample,
  title={From sample average treatment effect to population average treatment effect on the treated: combining experimental with observational studies to estimate population treatment effects},
  author={Hartman, Erin and Grieve, Richard and Ramsahai, Roland and Sekhon, Jasjeet S},
  journal={Journal of the Royal Statistical Society: Series A (Statistics in Society)},
  volume={178},
  number={3},
  pages={757--778},
  year={2015},
  publisher={Wiley Online Library}
}

@article{bartlett2006convexity,
  title={Convexity, classification, and risk bounds},
  author={Bartlett, Peter L and Jordan, Michael I and McAuliffe, Jon D},
  journal={Journal of the American Statistical Association},
  volume={101},
  number={473},
  pages={138--156},
  year={2006},
  publisher={Taylor \& Francis}
}

@inproceedings{collobert2006trading,
  title={Trading convexity for scalability},
  author={Collobert, Ronan and Sinz, Fabian and Weston, Jason and Bottou, L{\'e}on},
  booktitle={Proceedings of the 23rd international conference on Machine learning},
  pages={201--208},
  year={2006}
}

@article{nocedal1980updating,
  title={Updating quasi-Newton matrices with limited storage},
  author={Nocedal, Jorge},
  journal={Mathematics of computation},
  volume={35},
  number={151},
  pages={773--782},
  year={1980}
}

@article{zhou2017residual,
  title={Residual weighted learning for estimating individualized treatment rules},
  author={Zhou, Xin and Mayer-Hamblett, Nicole and Khan, Umer and Kosorok, Michael R},
  journal={Journal of the American Statistical Association},
  volume={112},
  number={517},
  pages={169--187},
  year={2017},
  publisher={Taylor \& Francis}
}

@article{le1997solving,
  title={Solving a class of linearly constrained indefinite quadratic problems by DC algorithms},
  author={Le Thi Hoai, An and Tao, Pham Dinh},
  journal={Journal of global optimization},
  volume={11},
  number={3},
  pages={253--285},
  year={1997},
  publisher={Springer}
}

@article{wu2007robust,
  title={Robust truncated hinge loss support vector machines},
  author={Wu, Yichao and Liu, Yufeng},
  journal={Journal of the American Statistical Association},
  volume={102},
  number={479},
  pages={974--983},
  year={2007},
  publisher={Taylor \& Francis}
}

@book{Vapnik1998,
  author = {Vapnik, Vladimir N.},
  publisher = {Wiley-Interscience},
  title = {Statistical Learning Theory},
  year = 1998
}

@inproceedings{kallus2017recursive,
  title={Recursive partitioning for personalization using observational data},
  author={Kallus, Nathan},
  booktitle={Proceedings of the 34th International Conference on Machine Learning-Volume 70},
  pages={1789--1798},
  year={2017},
  organization={JMLR. org}
}

@article{zhang2012robust,
  title={A robust method for estimating optimal treatment regimes},
  author={Zhang, Baqun and Tsiatis, Anastasios A and Laber, Eric B and Davidian, Marie},
  journal={Biometrics},
  volume={68},
  number={4},
  pages={1010--1018},
  year={2012},
  publisher={Wiley Online Library}
}

@article{zhang2012estimating,
  title={Estimating optimal treatment regimes from a classification perspective},
  author={Zhang, Baqun and Tsiatis, Anastasios A and Davidian, Marie and Zhang, Min and Laber, Eric},
  journal={Stat},
  volume={1},
  number={1},
  pages={103--114},
  year={2012},
  publisher={Wiley Online Library}
}

@article{rubin1974estimating,
  title={Estimating causal effects of treatments in randomized and nonrandomized studies.},
  author={Rubin, Donald B},
  journal={Journal of educational Psychology},
  volume={66},
  number={5},
  pages={688},
  year={1974},
  publisher={American Psychological Association}
}

@article{wang2020minimal,
  title={Minimal dispersion approximately balancing weights: asymptotic properties and practical considerations},
  author={Wang, Yixin and Zubizarreta, Jose R},
  journal={Biometrika},
  volume={107},
  number={1},
  pages={93--105},
  year={2020},
  publisher={Oxford University Press}
}

@article{zhao2012estimating,
  title={Estimating individualized treatment rules using outcome weighted learning},
  author={Zhao, Yingqi and Zeng, Donglin and Rush, A John and Kosorok, Michael R},
  journal={Journal of the American Statistical Association},
  volume={107},
  number={499},
  pages={1106--1118},
  year={2012},
  publisher={Taylor \& Francis}
}

\end{document}